\documentstyle[aps,prl,floats]{revtex}
\begin{document}
\input epsf.sty

\twocolumn[\hsize\textwidth\columnwidth\hsize\csname 
@twocolumnfalse\endcsname

\title{Ultraslow Electron Spin Dynamics in GaAs
Quantum Wells 
 Probed by Optically Pumped NMR}

\author{N. N. Kuzma$^{1}$, P. Khandelwal$^{1}$, S. E.
Barrett$^{*1}$,  L. N. Pfeiffer$^{2}$, K. W.
West$^{2}$}

\address{$^{*}$To whom correspondence should be
addressed.\\
 $^{1}$Department of Physics, Yale University, New Haven,
 CT, USA 06511 \\ $^{2}$Lucent Technologies Bell
Laboratories, Murray
 Hill, New Jersey 07974}

\date{\today}

\maketitle

\begin{abstract}  Optically pumped nuclear
magnetic resonance (OPNMR) measurements 
were performed in two different
electron-doped multiple quantum well samples
near the fractional quantum Hall effect ground 
state $\nu$=$\frac{1}{3}$.  
Below 0.5\,kelvin, the spectra provide 
evidence that spin-reversed charged excitations
of the $\nu$$=$$\frac{1}{3}$ ground state are
localized over the NMR time scale of
about 40 microseconds.
Furthermore, by varying NMR pulse
parameters, the electron spin temperature 
(as measured by the Knight shift)
 could be 
driven above the lattice temperature,
 which shows that 
 the value of the 
electron spin-lattice relaxation 
time $\tau_{1s}$ is between
100 microseconds  and  500 milliseconds
 at $\nu$=$\frac{1}{3}$.
\end{abstract}

\vskip 2pc ] 

\narrowtext

A two-dimensional electron system (2DES),  cooled
to extremely low temperatures in  a strong
magnetic field, exhibits many exotic phenomena, such as the
fractional quantum Hall effect (FQHE) \cite{tsui}.
  Transport 
and optical studies of the 2DES have shown that the low-energy
 physics in these extreme conditions is driven by
the electron-electron Coulomb interaction \cite{laughlin},
 but the challenge
of precisely describing the low-lying many-body states
that exist in a real 2DES remains formidable for both
theory and experiment\cite{chakraborty,prange,das sarma}.
Such 2DESs have been probed by 
 OPNMR\cite{opnmrchar,opnmrnu1,tycko}
  in the FQHE
 regime\cite{khandelwal}, which allows 
 the direct radio-frequency (rf)
detection of NMR signals from nuclei in electron-doped
GaAs quantum wells.  The $^{71}$Ga OPNMR spectra 
 reveal the 
local, time-averaged value of the electron
spin magnetization,
 $\langle S_{z}(\vec{R})\rangle$, thus leading to
insights
 about the many-electron states relevant for the FQHE.  
  
We report 
  evidence of ultraslow electron spin
dynamics near the most studied FQHE ground
 state, $\nu$=$\frac{1}{3}$, with characteristic
time scales exceeding $\sim$40$\mu$s below 0.5\,K.
 Although the samples are characterized by ``simple" NMR
parameters (that is,
isotropic  hyperfine coupling in an oriented 
single crystal), the OPNMR
spectra are complex because they can
  simultaneously exhibit inhomogeneous broadening
due to the quantum confinement
 of electrons within a well and motional narrowing
due to 
 delocalization of electrons along
 the well\cite{slichter}.  At low temperatures ($T$$\approx$0.5 K),
a  change
in the NMR linewidth is 
observed whenever spin-reversed electrons 
are present.  We attribute this striking 
behavior  to the localization of spin-reversed
electrons over the NMR time scale.  In addition, using
the Knight shift as a thermometer, we measured the 
increase of the electron spin temperature above the lattice
temperature when rf pulses were used to drive the system out of 
equilibrium.  These non-equilibrium measurements 
imply that the electron spin-lattice relaxation time
is $100\,\mu$s$<$$\tau_{1s}$$<$\,500\,ms for $T<$0.5\,K
at $\nu$=1/3, which appears to exceed all electronic time scales
 previously measured in semiconductors by at least a factor
of 1000.

Both of the multiple quantum well samples in this study
were grown by molecular beam epitaxy on semi-insulating
GaAs(001) substrates.
  Sample  40W contains 40 $300\,$\AA\space wide GaAs
wells that were separated
 by $3600\,$\AA\space wide Al$_{0.1}$Ga$_{0.9}$As
barriers.   Sample 10W  contains 10
$260\,$\AA\space wide wells that were separated by 
$3120\,$\AA\space wide 
 barriers.  Silicon delta-doping spikes located at  the
center of each barrier provided the electrons that were
confined in each GaAs well at low temperatures,
producing a 2DES with very high mobility 
($\mu>1.4\times10^{6}$~cm$^{2}$V$^{-1}$s$^{-1}$)\cite{pfeiffer}.
  The 2D
electron densities in each well were
$n_{\text{40W}}\,$=$\,6.69\,$$\times\,$$10^{10}\,\text{cm}^{-2}$
and
$n_{\text{10W}}\,$=$\,7.75\,$$\times\,$$10^{10}\,\text
{cm}^{-2}$\cite{khandelwal}.

 The low-temperature ($0.29\,$K$<$$\,T$$\,<\,$$1.5\,$K),
 high-field ($B_{\text{tot}}$=12 T)
 OPNMR measurements 
 were performed with an Oxford Instruments 
sorption-pumped~$^{3}$He cryostat that was mounted in a 
 Teslatron$^H$ superconducting magnet.  The
samples, about 4 mm by 6 mm by 0.5 mm, were in direct
contact with helium and were mounted on the platform of a
rotator assembly in the NMR probe. By tilting this
platform, we could vary the angle $\theta$
$(-60^\circ$$<$$\theta$$<$$60^\circ$, $\pm$$0.1^\circ$)
between the sample's growth axis $\vec{z^{\prime}}$
(perpendicular to the plane of the wells) and the applied
field $\vec B_{\text{tot}}$ (fixed along $\vec{z}$), thereby
changing the filling factor $\nu$=$(nhc)/(eB_{\bot})$ 
in situ (here
$B_{\bot}$$\equiv$$B_{\text{tot}}\!\cos\theta$), where 
$h$ is Planck's constant, $c$ is the speed of light,
and $e$ is the electron charge.
  For
 optical pumping, light from a Coherent 890
Ti:Sapphire laser was delivered into the cryostat 
through an
optical fiber\cite{heiman}, which was
 terminated by a collimating lens
and a polarizing assembly 22 cm above the
sample.  The light spot on the sample (5\,mm diameter,
812\,nm wavelength, left-circularly polarized,
 $\leq$10\,mW cm$^{-2}$)
was gated by a spectrometer-controlled 
room-temperature shutter.

For the OPNMR measurements, we used the 
 timing sequence
\mbox{SAT--$\tau_{L}$--$\tau_{D}$--DET}
\cite{opnmrchar,opnmrnu1,tycko,khandelwal};
 SAT denotes an rf pulse train that
destroys (saturates) the $^{71}$Ga nuclear polarization
throughout the sample, 
$\tau_{L}$ is light time, $\tau_{D}$ is dark time,
and DET denotes the detection period.
  During $\tau_{L}$ (30 to
90\,s), optical pumping of interband transitions
generated electrons and holes in the GaAs wells with
nonequilibrium spin polarizations, which then polarized
the nuclei in the wells through the hyperfine
interaction. The shutter was then closed to allow the
electrons to equilibrate with the $^{3}$He bath during
$\tau_{D}$ (typically 40\,s). The enhanced nuclear
polarization persisted until DET,
 whereupon
a single rf tipping pulse was applied to produce
a free induction decay signal, which we then acquired with
a home-built NMR spectrometer that was
based on a Tecmag Aries system.
A calibrated RuO$_{2}$ thermometer, in good
thermal contact with the sample, recorded the
temperature during signal acquisition.  

\begin{figure}
\centerline{\epsfxsize=3.55in\epsfbox{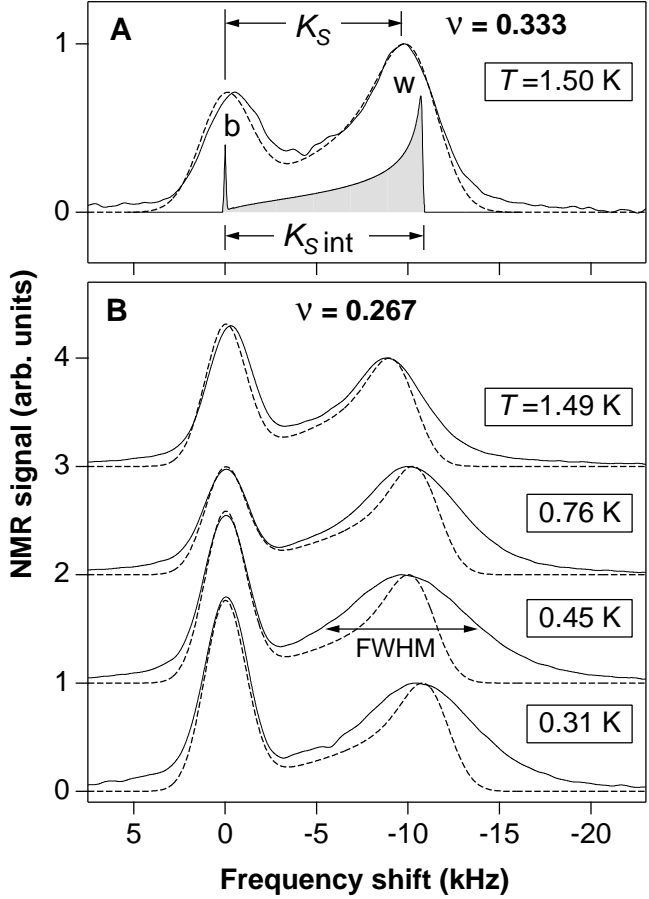}}
\caption{({\bf A}) The $^{71}$Ga OPNMR emission spectrum 
(solid line) of sample 10W
at $\nu$=$\frac{1}{3}$, taken at
 $\theta $=36.8$^\circ$ in
$B_{\text{tot}}$=12\,T. The frequency shift is
relative to $f_o$=155.93\,MHz.
The dashed line fit is obtained by
 broadening the intrinsic
line shape (hatched region).  Empirically,
$K_{S\text{int}}$=$K_S$$+$$1.1$$\times$$(1$$-$$
\exp(-K_S/2.0))$ (in kilohertz).  ({\bf B}) Temperature
 dependence of the
spectra at $\nu$=0.267 ($\theta $=0$^\circ$).  The FWHM 
 of the well resonance w peak is shown.
Arb. units, arbitrary units.}
\label{fig1}
\end{figure}

 A $^{71}$Ga OPNMR emission
spectrum at $\nu$=$\frac{1}{3}$ 
(Fig.\,\ref{fig1}A, solid line) exhibits a ``w" peak 
 that arises from nuclei in the GaAs quantum
well and a ``b" peak  that is due to nuclei in
the Al$_{0.1}$Ga$_{0.9}$As 
barriers\cite{opnmrnu1,tycko,khandelwal}.  The Fermi contact
hyperfine coupling between the spins of the 2DES
and nuclei in the well shifts the w peak  below
the b peak by K$_S$, which we define to be
 the Knight shift\cite{slichter,winter}.
The asymmetry of the well line shape has two origins:
(i) the quantum confinement within the well 
causes the electron density to vary across its width $w$
as $\rho$(z$^{\prime}$) $\approx \cos ^2$($\pi$z$^{\prime}$/$w$) 
for $|$z$^{\prime}|\leq$$w$/2\cite{weisbuch,fishman} 
 and (ii) the optical pumping
preferentially polarizes nuclei in the center of the well.
Taking these two effects into account, 
the intrinsic line shape (Fig.\,1A,
 hatched region) may be written as the sum of 
$I_{\text{w}}^{\text{int}}\!(\textstyle K_{S\text{int}},f)$
=$[f/(K_{S\text{int}}-f)]^{1/2}$ 
and $a_b \delta (0)$ for the unbroadened barrier 
signal.  Using a  3.5-kHz full width at half maximum (FWHM)
 Gaussian 
 $g(f)$ for  the nuclear dipolar broadening
\cite{slichter}, we arrive at a two-parameter fit (Figs. 1
and 2, dashed lines): 

\begin{displaymath}     I(f)=I_{\text{b}}+I_{\text{w}}=
a_{\text{b}}\,g(f)+\!\int^{\textstyle ^{K_{S\text{int}}}
}_{\textstyle _{0}}
\!\!\!\!\!\!\!\!\!\!\!\!\!\!df'  g(\,f\!-\!f'\,)\,
I_{\text{w}}^{\text{int}}\!(\textstyle
K_{S\text{int}},f')
\end{displaymath}

The first parameter, $a_{\text{b}}$, is the amplitude of the
barrier signal, which grows
during $\tau_{L}$ as the optically pumped 
nuclear magnetization diffuses out of the quantum well.
 The second parameter
extracted from the fit is the hyperfine shift
for nuclei in the center of the well, 
$K_{S\text{int}}(\nu,T)\,$
=${\cal P}(\nu,T)\:\frac {n}{w}\:(4.5\pm 0.2)$
$\times$$10^{-16}\,$kHz\,cm$^{3}$\cite{khandelwal}.
Thus, fits to OPNMR spectra at various $\nu$ and $T$
provide a direct measure of the electron spin polarization
${\cal P}(\nu,T) \equiv \frac{\langle S_{z}(\nu,T)\rangle}
{\text{max}\langle S_{z}\rangle}$
 in the quantum well.

This approach has  been used to map out
${\cal P}(\nu,T)$ in the vicinity of important
integer and 
fractional quantum Hall ground states of
$\nu$=1 and $\nu$=$\frac{1}{3}$.  The ${\cal P}(T)$ measurements
at fixed $\nu$ revealed the neutral spin flip excitations of
the fully polarized ground states.   Measurements
of ${\cal P}(\nu)$ provided the first evidence\cite{opnmrnu1,tycko}
of the existence of skyrmions\cite{sondhi,fertig}, which are
charged spin-texture excitations of the $\nu$=1 ground
state\cite{das sarma}. Recently,
 ${\cal P}(\nu)$ was found to drop on either
side of the $\nu$=$\frac{1}{3}$ ground state, which shows
that the charged excitations of this FQHE ground state
are partially spin reversed, even in a 
12-T field\cite{khandelwal}. 

In all of these earlier measurements, the OPNMR spectra were
well described by the dashed-line fits generated by our
model.
The  central assumption of this model is that the electron
spins are delocalized along the well, such that 
$\langle S_{z}(\nu,T)\rangle$ appears spatially
homogeneous  when averaged over the NMR time scale
($\sim\,$40$\,\mu$s)\cite{nmrtime}. In this limit, the
delocalization of the low density 2DES (there are
$\sim 10^{6}$ nuclei per electron in the well) produces 
a motional narrowing of $I_{\text{W}}^{\text{int}}$. 

However,  
low-temperature measurements at $\nu$=0.267 
 showed a 
crossover to more complicated line shapes (Fig.~1B). 
 Although the spectra were in reasonable
agreement with our model above 1\,K, the width of the  w
peak increased dramatically as the temperature was
lowered to $T$=0.45\,K and then decreased 
upon further lowering to $T$=0.31\,K.
This 
nonmonotonic temperature dependence 
 is reminiscent of the behavior seen in NMR
studies  of systems in which spectra are sensitive to
dynamical processes\cite{abragam},
 variously referred to as motional
narrowing, dynamical averaging,
 or chemical exchange\cite{slichter,kaplan,mehring}.
 In our experiment, the
nuclei were rigidly fixed in the lattice of a single
crystal, so the variation in the line shape shown in Fig.\,1B
 was a signature of electron spin localization, which
turned off the motional narrowing of the well resonance
as the temperature was lowered.

\begin{figure}
\centerline{\epsfxsize=3.55in\epsfbox{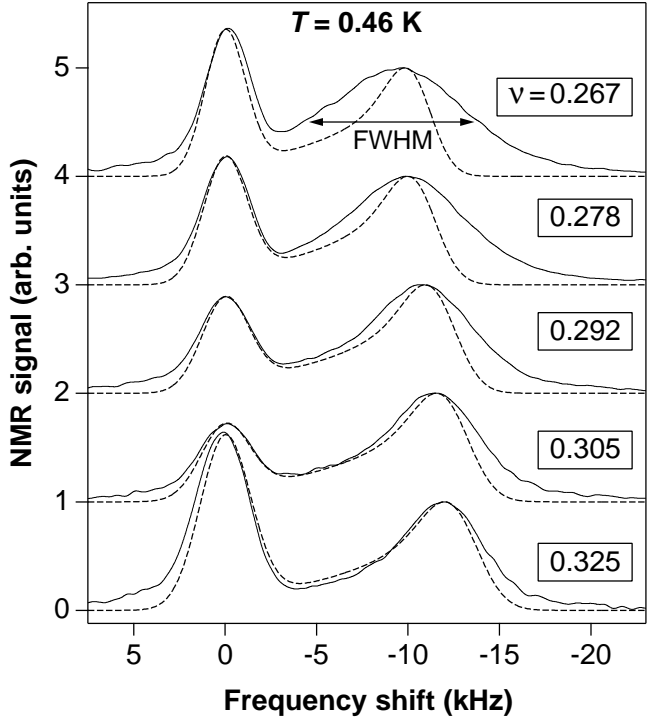}}
\caption{The $^{71}$Ga OPNMR spectra (solid lines) of sample 10W
at $T$=0.46\,K, for 0.267$\leq$$\nu$$<$1/3
 (0$^\circ$$\leq$$\theta$$<$36.8$^\circ$).
}
\label{fig2}
\end{figure}

 The extra broadening of the well resonance 
disappeared as the sample  was tilted from
 $\theta_{10\text{W}}$=0$^\circ$ ($\nu$=0.267)
 to 36.8$^\circ$ ($\nu=1/3$) (Fig.2),
despite a 10$\%$ increase in the dipolar 
broadening (the $^{75}$As nearest-neighbors
 of the $^{71}$Ga nuclei are at the ``magic
angle"\cite{mehring} when $\theta_{10\text{W}}$=0$^\circ$).
Furthermore,  there was a 
correspondence between the  decrease in 
the linewidth 
 and the  increase 
in $K_S$  as $\nu$$\rightarrow$1/3 (Fig.~3).
This anticorrelation strongly suggests that 
the behavior shown in Figs.~1 to 3 is due to
electron spin dynamics.

For a quantitative understanding of these phenomena, we must 
 consider the specific assumptions that
lead to 
$I_{\text{w}}^{\text{int}}\!(\textstyle K_{S\text{int}},f)$.
  Nuclei within
the well couple to the spins of the 2DES through the
isotropic Fermi contact
 interaction\cite{opnmrchar,opnmrnu1,tycko,khandelwal,slichter};
 thus, a nucleus at site
 $\vec{\text R}$$^{\prime}$ experiences a hyperfine
magnetic field 
\mbox{$\vec B$$^{e}$($\vec{\text R}$$^{
\prime}$)=($-16$$\pi$$\mu_{B}$/3)$\sum_{
\text j}$$\vec S_{\text j}$$\delta$($\vec{
\text r}_{\text j}$$-$$\vec{\text R}$$^{\prime}$)}, 
where $\mu_{B}$ is the Bohr magneton, $\vec S_{\text j}$
is  the spin of electron j, the summation is over all of
the conduction electrons within the well, and the delta
function picks out those electrons that  overlap with
the nucleus at $\vec{\text R}$$^{\prime}$. The average
projection of $\vec B$$^{e}$ along the applied field
$\vec B$$_{\text{tot}}$ may be written quite generally
as 
$\langle B$$_z^e$($\vec{\text
R}$$^{\prime}$,$\nu$,T)$\rangle $ = ($\frac {-8 \pi
\mu_{ B}}{3}$)($\frac {n}{w}$)($
|^{71}$u(0)$|^2$$|\chi$(Z$^{\prime}$)$|^2$$|\phi$(X$^{
\prime}$,Y$^{\prime}$)$|^2$)$\cal P$($\vec{\text R}$$^{
\prime}$,$\nu$,T). Here, the probability density of
finding electrons at a $^{71}$Ga site has been
factored into a term with the periodicity of the lattice
$|^{71}$u(0)$|^2$ and terms that vary slowly 
within a unit cell
 $|\chi$(Z$^{\prime}$)$|^2$$|\phi$(X$^{
\prime}$,Y$^{\prime}$)$|^2$\cite{winter}.
$\cal P$($\vec{\text R}$$^{\prime}$,$\nu$,T) is the
local  spin-polarization ($-1$$<$$\cal P$$<$1) of the electrons at 
$\vec{\text R}$$^{\prime}$. 
If we  assume that electrons are delocalized
along the well, then
the time-averaged values of $|\phi|$$^2$ and $\cal
P$ are spatially homogeneous.  In this
limit,  the local hyperfine frequency shift (taken to have
the sign of $\cal P$) is a
function of z$^{\prime}$ only, 
[f(z$^{\prime}$)=$-^{71}$$\gamma$$\langle
B$$_{z}^e$(z$^{
\prime}$,$\nu$,T)$\rangle $/2$\pi$$\approx$$\cos ^2$($\pi$z$^{
\prime}$/w)$K$$_{S\text{int}}$], so the general
expression for the well line shape is 
{$I_{\text{W}}^{\text{int}}$(f$^{\prime}$)=$\sum
_{\text{N-wells}}$
$\int$\mbox{d$^3$$r^{\prime}$}$\langle
$$^{71}$I$_z$($\text z$$^{\prime}$)$\rangle $$
\rho_{\text {nuclear}}$$\delta$(f$
^{\prime}$-f(z$^{\prime}$)).  We further assume that
wells are  identical
 and that the optical pumping gives rise to a
 nuclear polarization that varies across each well as
 $\langle$$^{71}$I$_z$($\text z$$^{\prime}$)$\rangle$
$\sim$ f(z$^{\prime}$), which leads to the form
$I_{\text{W}}^{\text{int}}$(f$^{\prime}$) = 
$[f^{\prime}/(K_{S\text{int}}-f^{\prime})]^{1/2}$
 (shown in Fig.\,1A).

\begin{figure}
\centerline{\epsfxsize=3.55in\epsfbox{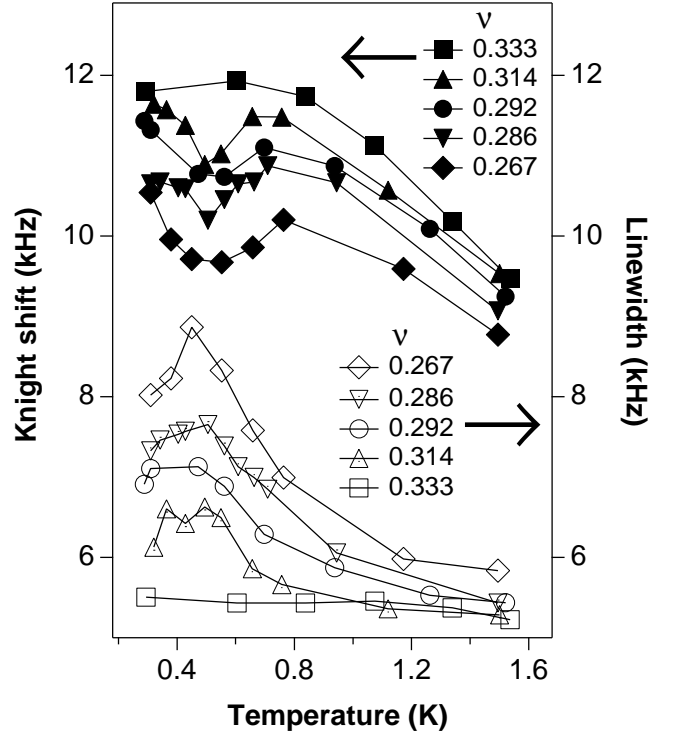}}
\caption{The temperature dependence of the Knight shift
(solid symbols) and the linewidth (open symbols) for
several filling factors 0.267$\leq$$\nu$$\leq$1/3 in
sample 10W. Lines are
to guide the eye.}
\label{fig3}
\end{figure}

The observed broadening of the well line shape beyond
the motionally narrowed limit implies that the
time-averaged value of $|\phi|$$^2$$\cal
P$ becomes spatially inhomogeneous.
Although the $|\phi$(X$^{\prime}$,Y$^{\prime}$)$|^2$ term
could become inhomogeneous if a pinned Wigner crystal
were to form, the corresponding increase in the 
linewidth (by orders of
magnitude) and the concomitant drop in the peak intensity 
are not observed.
Furthermore, variations in charge density 
along the well (for example, arising from fluctuations 
in the dopant layer) do not appear to explain either the 
magnitude of the effect\cite{density}
 or the nonmonotonic temperature
 dependence.  In constrast, the Knight shift data
show that the total spin polarization drops monotonically
below $\nu$=$\frac{1}{3}$, allowing the local spin
polarization $\cal P$($\vec{\text R}$$^{\prime}$)
to be spatially inhomogeneous.
Thus, we conclude that localization of spin-reversed regions
is responsible for the behavior shown in Figs. 1 to 3.

\begin{figure}
\centerline{\epsfxsize=3.55in\epsfbox{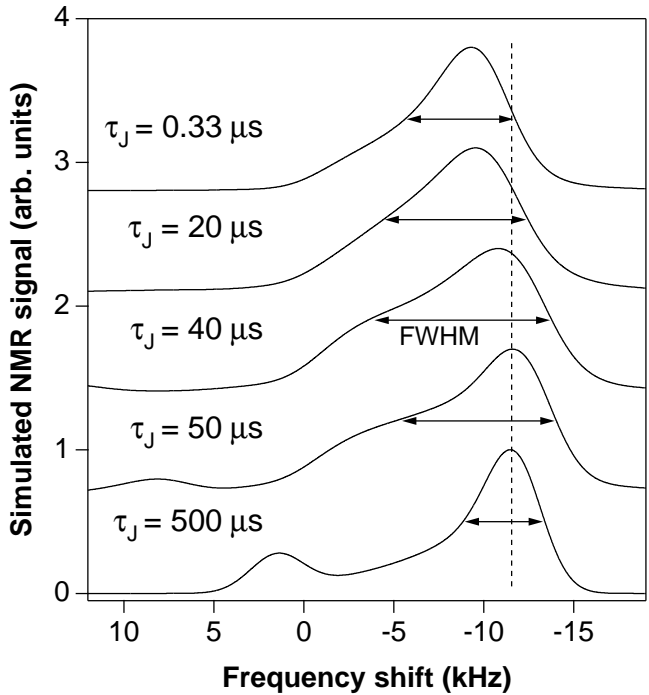}}
\caption{OPNMR spectra simulated with the model described 
in the text. 
 $K_{S\text{int}}$ is set to 12\,kHz for $\cal P$=1.
The barrier is suppressed ($a_b$=0) for clarity.}
\label{fig4}
\end{figure}

The time scale of the  spin localization  may be inferred
by simulating the observed line shapes.
In our model, for every point ($x',y'$)
along the plane of the well, the local polarization is either
up ($\cal P$=1) or down ($\cal P$=-0.15).  After
every jump time $\tau_{J}$, the local polarization 
instantaneously assumes either
 the up or down value with probability p$_{_{+}}$
or  (1-p$_{_{+}}$), respectively.
At all times, the ratio of up to down sites
is p$_{_{+}}$/(1-p$_{_{+}}$).   
 The 
simulated OPNMR spectra depend upon the value of $\tau_{J}$,
as is shown in Fig.4 for
the case p$_{_{+}}$=0.85.  The simulation  is in
reasonable agreement with the corresponding data from sample 10W.
When  $\tau_{J}$ is very fast, all nuclei see the same
  time-averaged  local polarization, which is equal to 
the total polarization
($\cal P_{\text{total}}$=0.828 at $\nu$=0.275 for our assumptions).  At the
other extreme ($\tau_{_{\text J}}$$\rightarrow$$\infty$),
 the motion is frozen out, and the single resonance
is split into up and  down
 lines, with  areas
proportional to p$_{_{+}}$ and
 (1-p$_{_{+}}$), respectively.  Even
within this simple model, the inhomogeneous
 breadth of the frozen
line shape (owing to the quantum confinement)
 leads to a nontrivial
evolution of the spectrum in the 
intermediate motion regime (for example,
a given value of  $\tau_{_{\text J}}$ 
might be simultaneously
``fast'' for nuclei at the edge of the well 
and ``slow'' for nuclei in
the center of the well).  In the intermediate motion
regime, the FWHM of the w peak goes through a maximum
when $\tau_{_{\text J}}$=40\,$\mu$s.  Although
 varying the parameters
p$_{_{+}}$ and $K_{S\text{int}}$ (over the range relevant for
samples 10W and 40W)  affects the extreme value of the FWHM,
 the characteristic $\tau_{_{\text J}}$ remains 
$\sim$40\,$\mu$s. 

 On the basis of this simple model,
the peaks in the FWHM at T$_{\text{loc}}$ $\approx$ 0.5\,K 
(Fig.\,3) reflect the localization temperature
of reversed spins,
such that they fail to cover the sample uniformly over
$\sim 40\,\mu$s.  The self-similar 
 curves in Fig.\,3 suggest that T$_{\text{loc}}$
 is not a strong function of the filling factor (or 
the density of reversed spins) for $\nu$$<$1/3.
Below 0.5\,K, the measured
 $K_S$($\nu$$<$$\frac{1}{3}$) 
increases toward $K_S$($\nu$=$\frac{1}{3}$),
 as seen in the model. However, even down to
 $T$=0.3\,K, the spectra do not appear to match
 the frozen limit of
our simulation. 
As $\nu$ is varied below
1/3, the trends in the  $K_S$ and
FWHM data (Fig.\,3) continue smoothly
through $\nu$=2/7 without interruption. High-field
 magnetotransport measurements on samples
taken from the same wafer as 10W show
much more structure, with
 well-developed minima in $\rho_{xx}$ at
 $\nu$=$\frac{1}{3}$, $\frac{2}{5}$,
$\frac{2}{7}$,
 and $\frac{1}{5}$ 
at $T\,$=$\,0.3$~K\cite{pfeiffer,willett}.

  Additional measurements
of the linewidth for 
$\nu$$>$1/3 in sample 10W 
were consistent with the above picture.
Measurements in sample 40W for $\nu$$\leq$1/3 were also
in qualitative agreement,  with one
 important quantitative
difference: $T_{loc}$ appeared to be
 shifted lower, so that only the high temperature
side of the peak in the FWHM was observed down to $T$
$\approx$ 0.3\,K. There was a similar sample variation in
the saturation temperature of $\cal P$$(\frac {1}{3})$, with
T$_{\text{10W}}^{\text{sat}}$$\approx$0.77\,K and
T$_{\text{40W}}^{\text{sat}}$$\approx$0.46\,K.  The
observed spectra contain more information than 
our simple simulation has revealed. A more sophisticated
model might include: (i) a detailed
structure for the reversed spin regions present below 
$\nu$=1/3, (ii) the 2D dynamics of these reversed
 spins, and (iii) the effects of thermally excited
spin flips, because
 $T_{\text{sat}}$ is not that much greater than
the $T_{\text{loc}}$.

\begin{figure}
\centerline{\epsfxsize=3.10in\epsfbox{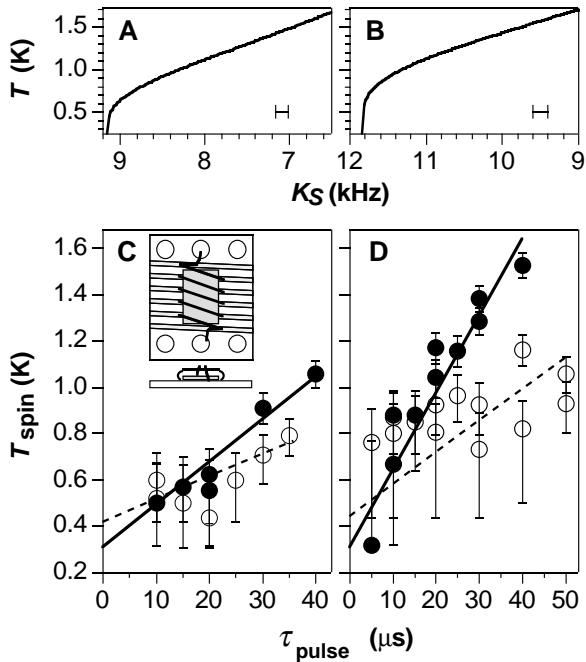}}
\caption{ T($K_S$) calibration curves based on
 the equilibrium 
$K_S$(T) data for ({\bf A}) sample 40W 
and ({\bf B}) sample 10W. Error bars for $K_S$ are shown.
 The dependence of
the effective spin temperature on the rf pulse length
($H_1$$\approx$7 Gauss) for ({\bf C}) sample 40W and
({\bf D}) sample 10W.  The intercept of the straight
line fit was constrained to
 be the lattice temperature: $T=0.31\,K$ (solid circles 10W and
40W),  $T=0.42\,K$ (open circles 40W), and
 $T=0.44\,K$ (open circles 10W). The inset (C) shows
the top (along $z'$) and the front (along the rotation
axis) views of   
the grooved sapphire platform holding a sample
in a 5-turn rf coil. }
\label{fig5}
\end{figure}

\begin{figure}
\centerline{\epsfxsize=3.10in\epsfbox{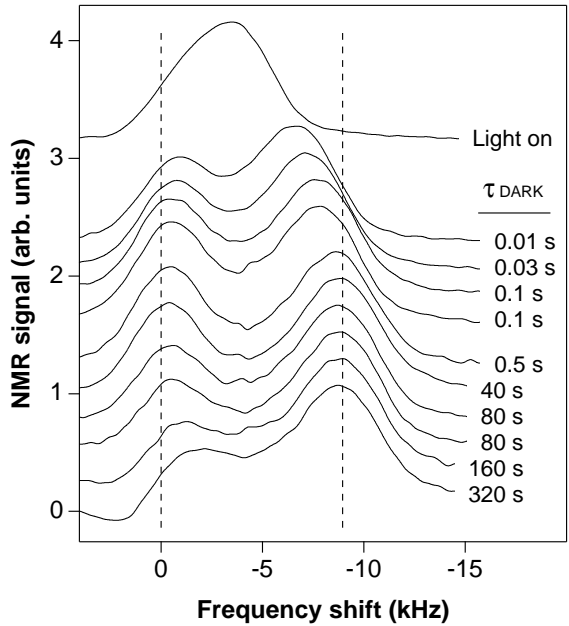}}
\caption{Spectra, which were acquired with sample 40W
at $T_{bath}$=0.45\,K, that show the evolution
of the line shape as a function of dark time $\tau_{D}$.}
\label{fig6}
\end{figure}

All of the results described thus
far were acquired by applying a weak rf
 tipping pulse 
long after optical pumping to probe 
the equilibrium properties of the
2DES.
 Non-equilibrium properties
of the electron spin system can be studied by
varying these parameters at low temperatures,
with a number of notable results at $\nu$=1/3.

The rf tipping pulse for the NMR experiment is produced by
a coil wrapped around the sample (Fig.\,\ref{fig5}C, inset),
which generates a linearly polarized
 (perpendicular to $\vec{z^{\prime}}$)
magnetic field of amplitude 2$\times $$H$$_{1}$ at 
$f_{o}$=155.93 MHz.  The equilibrium value of $K_S$(T) is
independent of the tipping pulse parameters for weak $H_1$
(that is, 
$H_1 \approx 5\,$Gauss, $\tau_{\text{pulse}}$=20\,$\mu$s).
However, if stronger pulses are used for T$<$0.5\,K, the measured
$K_S$ value drops sharply below the equilibrium value, even though
 the lattice temperature is unaffected by the pulses.  
The equilibrium measurements\cite{khandelwal}
 of $K_S$(T) (Fig.\,5, A and B),
can be used to convert the measured $K_S$
into an effective electron spin temperature $T_{\text{spin}}$;
 $T_{\text{spin}}$ rises linearly above
the lattice temperature $T$
 as the duration of the tipping pulse
$\tau_{\text{pulse}}$ increases, for $H_1\approx 7\,$G 
(Fig.\,5, C and D).
The increase of $T_{\text{spin}}$ drops off sharply with
increasing lattice temperature and is not observable for
$T$$>$0.5\,K.  Furthermore,
 the apparent heating depends strongly
on the alternating field strength and 
scales as $H_1^{\eta}$ (2$<$$\eta$$<$5), which rules out
 nuclear spins as the heat source,
 because their  tipping angle scales with
 $H_1$$\times$$\tau_{\text{pulse}}$.
Another possible mechanism, ohmic heating by eddy
currents, appears to be inconsistent with the strong
$T$ and $H_1$ dependence of the effect.
  Rather,
these data provide evidence for a direct coupling between
the rf pulse and the spins in the 2DES.  The mechanism for
this interaction in a clean system is not known, because
the applied rf photon energy is well below the electron
spin resonance  at $\sim$74 GHz.
Impurities in the
bulk or edge states may be involved in this process.

The nonequilibrium spectra 
remain motionally narrowed,
and appear to be
 indistinguishable from the corresponding equilibrium
spectra measured at a higher lattice temperature.
 Thus, the electron 
spin system achieves internal equilibrium before our
measurement, justifying our use of
 $T_{\text{spin}}$\cite{slichter}.
However, our measurement also shows that
$T_{\text{spin}}$ remains
 greater than $T$ long after the rf pulse is
turned off, which implies that the electron spin-lattice 
relaxation time
$\tau_{1s}$ is greater than
$100\,\mu$s  for T$<$0.5 K at $\nu$=$\frac{1}{3}$.

The evolution of the spectra as the  
dark time $\tau_D$  is increased  
provides an upper bound on $\tau_{1s}$ (Fig.\,6). 
The measured spectra are essentially independent of
$\tau_D$ after the first 0.5 s, which is consistent
with the equilibration time of the 
the laser-heated sample with 
the helium bath at 0.45\,K.  
 Combining these results, we find
$100\,\mu$s$<$$\tau_{1s}$$<$\,500\,ms  for $T<$0.5\,K
at $\nu$=1/3.
Although this value of $\tau_{1s}$ is at least a factor
of 1000 longer than recent measurements of the 
transverse relaxation time $\tau_{2}^{\ast}$ in 
bulk GaAs\cite{awshalom}, it is consistent with 
a previous theoretical prediction\cite{frenkel}
that had assumed conditions  similar to those
in our experiment.


\begin{references}
\bibitem{nmrtime} For nuclei in the center of
        the well, $\cal{P}$ can fluctuate between
        $-$1 and 1, resulting in local frequency
        variations of $\pm K_{S\text{int}}$.
        Thus the NMR time scale separating fast
        and slow fluctuations is $\sim$
        (2$K_{S\text{int}}$)$^{-1}$ $\approx$ 40 $\mu$s.
\bibitem{density}In simulations, 
        we found that an unreasonably large FWHM of 35\%
        was required for a gaussian distribution of electron
        densities along the well to explain the maximum
        NMR linewidth observed at $\nu$$=$0.267.
\end{references}
\end{document}